\documentclass{article}
\usepackage{spconf}
\usepackage[cmex10]{amsmath}
\usepackage{cite}
\usepackage{graphicx,epstopdf}
\usepackage{color}
\usepackage{times}
\usepackage{verbatim}
\usepackage{floatflt}
\usepackage{enumerate}
\usepackage{array}
\usepackage{multicol,afterpage,wrapfig}
\usepackage{tikz}
\usepackage{algorithm}
\usepackage[noend]{algpseudocode}
\makeatletter
\def\BState{\State\hskip-\ALG@thistlm}
\makeatother
\usepackage{amsmath,amsthm,amssymb,eucal}
\usepackage[utf8]{inputenc}
\usepackage{epsfig}
\usepackage{exscale}
\usepackage{latexsym}
\usepackage{verbatim}
\usepackage{amsfonts}
\usepackage{subfigure}
\usepackage{multirow}
\usepackage{cite}
\usepackage{balance}
\usepackage{color}
\usepackage{transparent}
\usepackage{import}
\usepackage{mathtools}
\usepackage{tikz}
\usetikzlibrary{automata,arrows,positioning,calc}
\usepackage{bbm}
\usepackage[printonlyused,withpage]{acronym}
\usepackage{tabu}
\usepackage{longtable}
\usepackage{mathtools}
\usepackage{stfloats}
\usepackage{psfrag}
\usepackage{float}
\usepackage{acronym}  
\usepackage{psfrag}

\acrodef{CCDF}{complementary cumulative distribution function}
\acrodef{CF}{characteristic function}
\acrodef{PPP}{Poisson point process}
\acrodef{RV}{random variable}
\acrodef{i.i.d.}{independent and identically distributed}
\acrodef{PDF}{probability distribution function}
\acrodef{CDF}{cumulative distribution function}
\acrodef{ch.f.}{characteristic function}
\acrodef{AWGN}{additive white Gaussian noise}
\acrodef{SNR}{signal-to-noise ratio}
\acrodef{LRT}{likelihood ratio test}
\acrodef{DRT}{distance ratio test}
\acrodef{GLRT}{generalized likelihood ratio test}
\acrodef{CRLB}{Cram\'{e}r-Rao lower bound}
\acrodef{CRB}{Cram\'{e}r-Rao bound}
\acrodef{ZZLB}{Ziv-Zakai lower bound}
\acrodef{ZZB}{Ziv-Zakai bound}
\acrodef{LOS}{line-of-sight}
\acrodef{ToF}{time-of-flight}
\acrodef{NLOS}{non-line-of-sight}
\acrodef{GDOP}{geometric dilution of precision}
\acrodef{GPS}{Global Positioning System}
\acrodef{FIM}{Fisher information matrix}
\acrodef{PEB}{position error bound}
\acrodef{SPEB}{squared position error bound}
\acrodef{TOA}{time-of-arrival}
\acrodef{TOF}{time-of-flight}
\acrodef{WSN}{wireless sensor network}
\acrodef{MAC}{medium access control}
\acrodef{RSS}{received signal strength}
\acrodef{WAF}{wall attenuation factor}
\acrodef{TDOA}{time difference-of-arrival}
\acrodef{RF}{radiofrequency}
\acrodef{RTT}{round-trip time}
\acrodef{AOA}{angle-of-arrival}
\acrodef{MF}{matched filter}
\acrodef{ED}{energy detector}
\acrodef{ML}{maximum likelihood}
\acrodef{MSE}{mean-square error}
\acrodef{RMSE}{root-mean-square error}
\acrodef{LEO}{localization error outage}
\acrodef{ppm}{part-per-million}
\acrodef{ACK}{acknowledge}
\acrodef{UWB}{Ultrawide bandwidth}
\acrodef{TNR}{threshold-to-noise ratio}
\acrodef{LS}{least squares}
\acrodef{IR-UWB}{impulse radio UWB}
\acrodef{FCC}{Federal Communications Commission}
\acrodef{TH}{time-hopping}
\acrodef{PPM}{pulse position modulation}
\acrodef{MUI}{multi-user interference}
\acrodef{PDP}{power delay profile}
\acrodef{BPZF}{band-pass zonal filter}
\acrodef{SIR}{signal-to-interference ratio}
\acrodef{SINR}{signal-to-interference-plus-noise ratio}
\acrodef{RFID}{radio frequency identification}
\acrodef{WPAN}{wireless personal area network}
\acrodef{WWB}{Weiss-Weinstein bound}
\acrodef{DP}{direct path}
\acrodef{MF}{matched filter}
\acrodef{MMSE}{minimum-mean-square-error}
\acrodef{SBS}{serial backward search}
\acrodef{SBSMC}{serial backward search for multiple clusters}
\acrodef{NBI}{narrowband interference}
\acrodef{WBI}{wideband interference}
\acrodef{INR}{interference-to-noise ratio}
\acrodef{CR}{channel response}
\acrodef{CIR}{channel impulse response}
\acrodef{CR}{channel  response}
\acrodef{RADAR}{radar}
\acrodef{MUR}{Multistatic radar}
\acrodef{JBSF}{jump back and search forward}
\acrodef{HDSA}{high-definition situation-aware}
\acrodef{RRC}{root raised cosine}
\acrodef{ST}{simple thresholding}
\acrodef{BTB}{Bellini-Tartara bound}
\acrodef{P-Max}{$P$-Max}  
\acrodef{MIMO}{multiple-input multiple-output}
\acrodef{MAP}{maximum a posteriori}
\acrodef{FG}{factor graph}
\acrodef{OP}{outage probability}
\acrodef{WED}{wall extra delay}
\acrodef{RMS}{root mean square}
\acrodef{SPAWN}{sum-product algorithm over a wireless network}
\acrodef{MDD}{minimum distance distribution}
\acrodef{MAP}{maximum a posteriori probability}
\acrodef{SAP}{small cell access point}
\acrodef{UE}{user equipment}
\acrodef{MBS}{macro cell base station}
\acrodef{UER}{\ac{UE} Relay}
\acrodef{D2D}{device-to-device}
\acrodef{MBS}{macro base station}
\acrodef{CSI}{channel state information}
\acrodef{OGR}{outage guard region}
\acrodef{FUR}{feasible UER region}
\acrodef{EHR}{energy harvesting region}
\acrodef{EH}{energy harvesting}
\acrodef{D2D-EHSN}{D2D communication provided \ac{EH} small cell network}
\acrodef{D2D-EHHN}{D2D communication provided \ac{EH} heterogeneous network}
\acrodef{3GPP}{3rd Generation Partnership Project}
\acrodef{BS}{base station}
\acrodef{DF}{decode and forward}
\acrodef{CCDF}{complementary cumulative distribution function}
\acrodef{ZF}{zero forcing}
\acrodef{RZF}{regularized zero forcing}
\acrodef{WLLN}{weak law of large number}
\acrodef{SLLN}{strong law of large numbers}
\acrodef{TDD}{Time-division duplex}
\acrodef{EE}{energy efficiency} 
\usepackage{color}
\usepackage{dsfont}
\usepackage{bbm}

\DeclareMathAlphabet{\mathsf}{OML}{cmbr}{m}{it}

\newcommand{\bd}{\begin{description}}
\newcommand{\ed}{\end{description}}
\newcommand{\be}{\begin{enumerate}}
\newcommand{\ee}{\end{enumerate}}
\newcommand{\bi}{\begin{itemize}}
\newcommand{\ei}{\end{itemize}}
\newcommand{\bl}{\begin{list}}
\newcommand{\el}{\end{list}}
\newcommand{\bt}{\begin{tabbing}}
\newcommand{\et}{\end{tabbing}}

  \graphicspath{{../Figures/}}
  \DeclareGraphicsExtensions{.eps}

\usepackage{amssymb}
\usepackage{amsthm}

\title{ Age-Based Scheduling Policy for Federated Learning \\ in Mobile Edge Networks }
\name{Howard H. Yang$^\dagger$, Ahmed Arafa$^\ast$, Tony Q. S. Quek$^\dagger$, and H. Vincent Poor$\ddagger$
}
\address{ $^\dagger$ Singapore University of Technology and Design, Singapore 487372 \\
$\ast$ University of North Carolina at Charlotte, NC 28223, USA \\
$\ddagger$ Princeton University, NJ 08544, USA}
				
\begin{document}
%

\maketitle

\begin{abstract}
Federated learning (FL) is a machine learning model that preserves data privacy in the training process.
Specifically, FL brings the model directly to the user equipments (UEs) for local training, where an edge server periodically collects the trained parameters to produce an improved model and sends it back to the UEs.
However, since communication usually occurs through a limited spectrum, only a portion of the UEs can update their parameters upon each global aggregation.
As such, new scheduling algorithms have to be engineered to facilitate the full implementation of FL.
In this paper, based on a metric termed the age of update (AoU), we propose a scheduling policy by jointly accounting for the staleness of the received parameters and the instantaneous channel qualities to improve the running efficiency of FL.
The proposed algorithm has low complexity and its effectiveness is demonstrated by Monte Carol simulations.
\end{abstract}


\section{Introduction}
Due to the sheer volume of data generated by the end-user devices, as well as
the increasing concerns about sharing private information, a new machine
learning model, namely the federated learning (FL) \cite{KonMcMBren:16,MaMMooRam:16,LinHanMao:18,SmiForMa:18}, has emerged from the intersection of
artificial intelligence and edge computing \cite{ZhaFenYan:19,WanTuoSal:19JSAC,LetCheShi:19}.
In stark contrast to the conventional machine learning methods that run in a data center, FL usually operates at the network edge and brings the models directly to the devices for training, where only the resultant parameters shall be sent to the edge servers that reside in an access point (AP). This salient feature of \textit{on-device training} brings along great advantages of eliminating the large communication overheads as well as preserving data privacy, and hence making FL particularly relevant for mobile applications \cite{WanTuoSal:19JSAC,LetCheShi:19,CheYanSaa:19,DuYanHua:19,AmiGun:19}. However, as the AP needs to link a large number of user equipments (UEs) over a limited spectrum, only a portion of the UEs can be selected to access the radio channel and send their trained updates in each global aggregation \cite{MaoYouZha:17,ZhaFenYan:19}.
To this end, the stragglers issue is more crucial for FL than for conventional trainings run in data centers \cite{LinHanMao:18,SmiForMa:18,HaZhaSim:19}.

Recognizing such criticality, a host of studies have been carried out and resulted in various scheduling protocols for FL, ranging from minimizing the transmission latency \cite{NisYon:19}, maximizing the spectral utility \cite{YanJiaShi:18}, to opportunistically alternating between selecting the UEs with advantageous and disadvantageous channel conditions \cite{ZhuWanHua:18}.
Even though positive gains have been demonstrated, these works put the main focus on exploiting the spectral resources so as to maximize the number of updates collectible by the AP in each round of global communication but ignore the staleness of these updates. As pointed out by \cite{DaiZhoDon:19}, the staleness of updates has a significant impact on the convergence rate of distributed machine learning models such as FL.
Therefore, it is reasonable to expect a scheduling algorithm that accounts for both the staleness and the communication quality can accelerate the convergence of FL in a mobile edge network.
On this purpose, we introduce a metric, referred to as the age-of-update (AoU){\footnote{Such age metric has been previously used, mainly, in the context of networking and has been shown to improve the notion of data freshness in various applications, see, e.g., the original treatment in \cite{KauYatGru:12}, and the survey in \cite{KosPapAng:17}.}}, to measure the staleness associated with each update and based on that, devise a scheduling protocol that allows the AP to collect timely updates from the UEs for FL training.
The algorithm is shown to possess low complexity, and its effectiveness is amply illustrated via Monte Carlo simulations.

\section{System Model}\label{sec:model}
\begin{figure}[t!]
  \centering{}

    {\includegraphics[width=0.85\columnwidth]{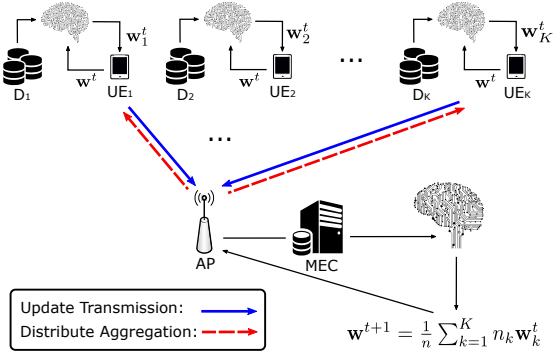}}

  \caption{ An illustration of the employed network model.
}
  \label{fig:FL_WN}
\end{figure}

Let us consider a mobile edge network that consists of one AP and $K$ user equipments (UEs), as depicted in Fig.~\ref{fig:FL_WN}, all are capable of performing local computing.
For a generic UE $k$, we consider that it is equipped with single antenna and has local data set $\mathcal{D}_k = \{ \mathbf{x}_i \in \mathbb{R}^d, y_i \in \mathbb{R} \}_{i=1}^{n_k}$ with size $n_k = |\mathcal{D}_k|$, where $|\cdot|$ denotes the cardinality.
In this network, the UEs communicate with the AP through a common spectrum, which is divided into $N$ equal-length subchannels. Here, we assume $N < K$.
The AP can possibly assign different UEs a different number of subchannels.
We adopt a block-fading propagation model, where the channels between any pair of antennas are assumed independently and identically distributed (i.i.d.) and quasi-static, i.e., the channel is constant during one transmission block, and varies independently from block to block.

The goal of the AP is to learn a statistical model over data that reside on the $K$ associated UEs. Specifically, the AP needs to fit a vector $\mathbf{w} \in \mathbb{R}^d$ so as to minimize a particular loss function. Formally, such a task can be expressed as follows:
\begin{align}\label{equ:Global_Obj_Func}
\min_{ \mathbf{w} \in \mathbb{R}^d } \Big\{ P( \mathbf{w} ) = \frac{1}{n} \sum_{ i=1 }^{n} \ell( \mathbf{w}; \mathbf{x}_i, y_i  ) + \xi r(\mathbf{w})  \Big\}
\end{align}
where $n = \sum_{ i=1 }^K n_i$ is the size of the whole data set, $\xi$ is the regularizing parameter and $r(\mathbf{w})$ a deterministic penalty function.
The function $\ell(\cdot)$ represents the loss function.
Several examples of loss functions used in popular machine learning models are summarized in \cite{YanLiuQue:19}.

Due to concerns of data privacy, the raw data set is generally not available at the AP and hence \eqref{equ:Global_Obj_Func} shall be solved using FL.
In particular, the training procedure (cf. Algorithm~3) comprises three major steps: $i$) the UEs conduct local computing aimed to solve \eqref{equ:Global_Obj_Func} by referring to the data resided on device and a downloaded global model, and send the resultant parameters to the AP, $ii$) the AP aggregates the received updates to produce an improved global model, and $iii$) the new model is sent back
to UEs, and the process is repeated.
After a sufficient amount of training and update exchanges, usually termed \textit{communication rounds}, between the AP and UEs, the objective function \eqref{equ:Global_Obj_Func} can converge to the global optimal.
Nevertheless, because the wireless medium is resource-constrained, the AP can only select a subgroup of UEs for parameter updates in each communication round, and the choice of selection can largely affect the convergence rate of FL \cite{YanLiuQue:19}.
In that respect, it languishes a dire need for appropriate scheduling algorithms.

\section{Scheduling Policy Design}
In this section,
we elaborate the concept of AoU, and detail the development of a scheduling policy based on this quantity.
\begin{figure}[t!]
  \centering{}
    {\includegraphics[width=0.95\columnwidth]{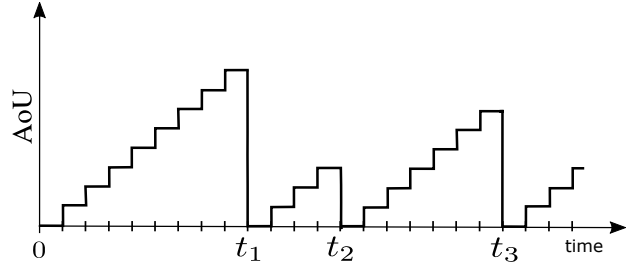}}
  \caption{ An example of the AoU evolution in time, where the transmissions take place at time stamps $t_1$, $t_2$, and $t_3$. }
  \label{fig:AoI}
\end{figure}
\subsection{ Design Metric }
In order to quantify the staleness of each update, we leverage the concept of information freshness \cite{KauYatGru:12} and define a metric termed Age-of-Update (AoU).
For a generic UE $k$, its AoU evolves as follows:
\begin{align}\label{equ:AgeProcess}
T_k[t+1] = ( T_k[t] + 1 ) ( 1 - S_k[t] ),~~ S_k[t] \in \{ 0, 1 \}
\end{align}
where $T_k[0] = 0$, and $S_k[t]$ takes value 1 if UE $k$ is selected by the AP for update during communication round $t$, and takes value 0 otherwise.
It is important to note that the AoU of UE $k$ measures the time elapsed since its latest update is received by the AP, and hence a larger AoU value indicates a higher degree of staleness associated with that UE.

As illustrated in Fig.~\ref{fig:AoI}, the evolution of AoU is affected by the scheduling algorithm.
Moreover, as the UEs are usually situated at various geographic locations, their AoU will progress differently due to their distinct communication conditions. In this regard, the different AoUs can be utilized as a reference to prioritize the channel access for the UEs.

\subsection{ Problem Formulation }
Using the notion of AoU, we devise our scheduling protocol such that the AP is able to keep the collection of all the updates ``as fresh as possible" by minimizing the sum of certain functions of the AoUs in each communication round. To be concrete, upon the global aggregation at round $t$, the AP selects the set of UEs for parameter update, denoted by $\mathbf{S}[t] = \{ S_1[t], S_2[t], ..., S_K[t] \}$, according to the following criteria:
\newline
\begin{subequations}\label{equ:Obj_Func}
\begin{align}
&\min_{ \mathbf{S}[t], \mathbf{P} } ~ \sum_{ k =1 }^K f_{\alpha}(\, T_k[t] \,) (\, 1 - S_k[t] \,)\\ \label{equ:PrimRst_Stb}
&~~\mathrm{s.t.}~ \left( \sum_{ n \in \mathcal{C}_k } \frac{1}{2} \log( 1 + G_{k,n} P_{k,n} ) - R_{ \mathrm{s} } \right) S_k[t] \geq 0 \\ \label{equ:PowCstr}
&~ \qquad \sum_{ n \in \mathcal{C}_k } P_{k,n} \leq P_{ \mathrm{TX} } \\
&~\, \qquad S_{k}[t] \in \{ 0, 1 \}, ~~~\forall \, k \in \{ 1, ..., K \} \\ \label{equ:ChlCstr}
&~\, \qquad \mathcal{C}_{i} \cap \mathcal{C}_{j} \!=\! \phi, ~ \mathrm{if}~ i \neq j ~  \mathrm{and} \, \cup_{ k = 1 }^K \mathcal{C}_k \! \subseteq \! \{ 1, ..., N \}
\end{align}
\end{subequations}
where $G_{k,n}$ is the gain attainable by UE $k$ on the $n$-th subchannel, $P_{k,n}$ the correspondingly injected power, and $\mathcal{C}_k$ is the set of subchannels assigned to UE $k$. Because the transmission of updates need to be finished within a certain time, the selected UEs should have their rates individually exceeding a threshold $R_{ \mathrm{s} }$, representing a required quality-of-service.
Moreover, the UEs are subject to a maximum transmit power through constraint \eqref{equ:PowCstr}, while the constraint \eqref{equ:ChlCstr} stipulates that the UEs shall follow an orthogonal channel access methodology to avoid interfering with each other.
The function $f_{\alpha}(\cdot)$ in \eqref{equ:Obj_Func} represents the \textit{sensitivity} of the edge server to the staleness of local updates. In this paper, we set the function as follows \cite{Atk:70}:
  \begin{align}\label{equ:f_alpha}
  f_{\alpha}(x) = \left \{
  \begin{tabular}{cc}
  \!\! $\frac{ x^{ 1 - \alpha } }{ 1-\alpha  } $, & $\mathrm{if}~\alpha \neq 1$,   \\
  \!\!\!\!  $\log( 1 + x )$, & $\mathrm{otherwise}$.
  \end{tabular}
  \right.
  \end{align}
The essence of the above function is to ensure a fairness treatment among the UEs based on their AoUs.

Two special cases are noteworthy: $i$) if $f_\alpha(x)$ is set as a constant, then solving problem~\eqref{equ:Obj_Func} is equivalent to packing the maximum number of UEs into the spectrum during each communication round \cite{YanJiaShi:18},
and $ii$) when the communication is conducted under an ideal environment, i.e., if $R_{\mathrm{s}} = 0$, solutions of the optimization problem~\eqref{equ:Obj_Func} lead to an $N$-Round-Robin policy \cite{YanLiuQue:19}.
Therefore, in the general scenario, the design problem given by \eqref{equ:Obj_Func} balances the tradeoff between fairness and quality on the radio channel access. It is also worthwhile to note that the design per problem \eqref{equ:Obj_Func} is not only suitable for scheduling updates in FL, but can also be leveraged in the scheduling of wireless traffic \cite{CaoLi:01}.

\begin{algorithm}[t!]
\caption{ Candidate UE List \& Subchannel Allocation }
\begin{algorithmic} \label{Alg:WaterFilling}
\State \textbf{Input:} The set of currently available subchannels $\bar{\mathcal{N}} \subseteq \mathcal{N}$
\State \textbf{Initialize:} $\mathcal{I}_{\mathrm{s}} = \phi$ and $\mathcal{C}_k = \phi, \forall k \in \{ 1, 2, ..., K \}$
\For { each UE $k \in \{ 1, 2, ..., K \}$ }
\State set $n_k^* = 1$ and order the channel gains as $G_{ k, \tau(1) } \geq G_{ k, \tau(2) } \geq ... \geq G_{ k, \tau( \bar{N} ) } $ where $\tau(n) \in \bar{\mathcal{N}}$
\While{$n_k^* \leq \bar{N}$}
\State Compute the following power updates:
\begin{align}
P_{k, \tau(n)} \!=\! \left \{
  \begin{tabular}{cc}
  \!\!\!\!\! $ \frac{ P(n_k^*) }{ n^*_k } + \frac{1}{ G_{ k, n_k^* } } - \frac{ 1 }{ G_{ k, \tau(n) } } $, &\!\!\!\! $ 1 \!\leq\! n \!\leq\! n^*_k $,   \\
  \!\!\!\!\!  $0$, &\!\!\!\! $ n^*_k \!<\! n \!\leq\! \bar{N} $
  \end{tabular}
  \right.
\end{align}
in which $P(n_k^*)$ is given by
\begin{align}
P(n_k^*) = \left[ P_{ \mathrm{TX} } - \sum_{ n=1 }^{ n_k^* - 1 } \left( \, \frac{ 1 }{ G_{ k, n_k^* } }  - \frac{ 1 }{ G_{ k, \tau(n) } } \, \right) \right]^{+}
\end{align}
where $[\cdot]^+ = \max(\cdot, 0)$
\If {$\sum_{ n = 1 }^{ n^*_k } \log( 1 + G_{ k, \tau(n) } P_{ k, \tau(n) } ) \geq R_{\mathrm{s}}$ }
\State Update $\mathcal{I}_{\mathrm{s}} = \mathcal{I}_{ \mathrm{s} } \cup \{ k \} $ and $\mathcal{C}_k = \cup_{n=1}^{n^*_k} \{ \tau(n) \}$
\State  Set $n_k^* = \bar{N} + 1$
\Else
\State Update $n_k^* = n_k^* + 1$
\EndIf
\EndWhile
\EndFor
\State \textbf{Output:} $\mathcal{I}_{\mathrm{s}}, \{ \mathcal{C}_k, k \in \mathcal{I}_{ \mathrm{s} } \}$
\end{algorithmic}
\end{algorithm}
\subsection{Algorithm}
The problem given by \eqref{equ:Obj_Func} is a combinatorial problem with mixed integers that does not possess low complexity solutions in general. Here, we tackle the problem via a greedy algorithm.
The approach is constituted of two major steps.

\begin{algorithm}[t!]
\caption{ Age-Based Scheduling (ABS) }
\begin{algorithmic} \label{Alg:Asgn_FL}
\State \textbf{Input:} $\mathcal{I}_{ \mathrm{s} }$, $\{ \mathcal{C}_k, k \in \mathcal{I}_{ \mathrm{s} } \}$, and $\mathbf{T}[t] = \{ T_k[t], k \in \mathcal{I}_{ \mathrm{s} } \}$
\State \textbf{Initialize:} $S_{k}[t] = 0, k \in \{ 1, 2, ..., K \}$
\If{ $\mathcal{I}_{ \mathrm{s} } \neq \phi$ }
\State Assign $S_{ k^* }[t] = 1$, in which the index $k^*$ is chosen as $k^* = \arg\max_{ k \in \mathcal{I}_{ \mathrm{s} } } f_\alpha( T_k[t] )/\Vert \mathcal{C}_k \Vert_0$
\State Update the available subchannels as $\bar{\mathcal{N}} = \bar{\mathcal{N}} \setminus \mathcal{C}_{k^*}$
\State Call Algorithm~1 with updated input $\bar{\mathcal{N}}$ and obtain the new outputs $\mathcal{I}_{\mathrm{s}}, \{ \mathcal{C}_k, k \in \mathcal{I}_{ \mathrm{s} } \}$
\State Call Algorithm~2 with updated inputs $\mathcal{I}_{\mathrm{s}}, \{ \mathcal{C}_k, k \in \mathcal{I}_{ \mathrm{s} } \}$
\EndIf
\State \textbf{Output:} $\mathbf{S}[t] = \{ S_1[t], S_2[t], ..., S_K[t] \}$
\end{algorithmic}
\end{algorithm}

First, amongst all the UEs, the AP needs to form a candidate list $\mathcal{I}_{\mathrm{s}}$ for further selection, by choosing those UEs that are able to achieve the targeted rate $R_{ \mathrm{s} }$ under the currently available spectrum budget. Moreover, each UE in the candidate list shall occupy the least amount of subchannels.
As such, a candidate UE $k$ needs to satisfy the following:
\begin{subequations} \label{equ:MinSubChnl}
\begin{align}
&\min_{ \mathcal{C}_k, \mathbf{P}_k } ~ \Vert \mathcal{C}_k \Vert_0 \\
&~~\mathrm{s.t.}~~ \sum_{ n \in \mathcal{C}_k } \frac{1}{2} \log( 1 + G_{k,n} P_{k,n} ) \geq R_{ \mathrm{s} } , \\ \label{equ:Rst_Stb}
& ~~\qquad \sum_{ n \in \mathcal{C}_k } P_{k,n} \leq P_{ \mathrm{TX} }
\end{align}
\end{subequations}
where $\Vert \cdot \Vert_0$ is the L-0 norm \cite{CanRomTao:06} and $\mathbf{P}_k = \{ P_{k,n}, n \in \mathcal{C}_k \}$  is the transmit power on the allotted subchannels.
For each selectioin of $\mathcal{C}_k$, subproblem \eqref{equ:MinSubChnl} can be solved by a water-filling algorithm, and the details of forming the candidate list is summarized in Algorithm~1.

Next, from the set of candidate UEs, the AP shall select the UEs that have relatively large AoU while requiring a relatively small amount of subchannels for the transmissions. Formally, the AP will select UE $i$ instead of UE $j$ if the following holds:
\begin{align}
f_\alpha( T_i[t] )/ \Vert \mathcal{C}_i \Vert_0 > f_\alpha(T_j[t]) / \Vert \mathcal{C}_j \Vert_0.
\end{align}
The above criteria brings us to the age-based scheduling (ABS) policy, summarized in Algorithm~2. Note that the ABS adopts a recursive search for UEs, since the occupation of a particular subchannel can incur other UEs in the candidate list to violate their rate constraint and hence get removed. Therefore, the candidate list $\mathcal{I}_{\mathrm{s}}$ needs to be recalculated via Algorithm~1 whenever a UE is selected.
Nevertheless, as the UE selection and subchannel assignments are conducted in a greedy manner, the algorithm has relatively low complexity.

Finally, we present the FL approach to solving \eqref{equ:Global_Obj_Func} in the setting of a mobile edge network, as per Algorithm~3.

\begin{algorithm}[t!]
\caption{ Wireless Federated Learning }
\begin{algorithmic} \label{Alg:Gen_FL}
\State \textbf{Parameters:} $\tau$ = number of local steps per communication round, $\eta$ = step size for local learning
\State \textbf{Initialize:} The AP assigns ${T}_k[0] = 0, \forall k \in \{1,...,K\}$ and randomly initializes $\mathbf{w}^0 \in \mathbb{R}^d$
\For { $t = 0, 1, 2, ..., T-1$ }
\State The AP runs ABS (Algorithm~2) to get $\mathbf{S}[t]$, and sends the parameter $\mathbf{w}^t$ to all the UEs in the updating list
\For { each UE with ${S}_k[t]=1$ }
\State Initialize $\mathbf{w}_k^t = \mathbf{w}^t$
\For { $s$ = 1 to $\tau$ }
\State Sample $i \in \mathcal{D}_k$ uniformly at random, and update the local parameter $\mathbf{w}^t_k$ as follows:
\begin{align}
\mathbf{w}_k^t \!=\! \mathbf{w}_k^t - \eta \big(\, \nabla \ell(\mathbf{w}_k^t; \mathbf{x}_i, y_i) + \xi \nabla r(\mathbf{w}_k^t) \, \big)
\end{align}
\EndFor
\State The UE updates the parameter $\mathbf{w}_k^t$ to the AP
\EndFor
\State The AP collects the trained weights $\{ \mathbf{w}^t_k \}_{ k \in \mathbf{S}[t] }$, and updates the AoUs $\mathbf{T}[t]$ and the global parameter $\mathbf{w}^{t+1}$ respectively as follows:
\begin{align}
&\mathbf{T}[t+1] = (\mathbf{T}[t] + \mathbf{1}) \odot ( \mathbf{1} - \mathbf{S}[t] ), \\
&\mathbf{w}^{t+1} = \frac{ \sum_{ k \in \mathbf{S}[t] } |\mathcal{D}_k| \mathbf{w}^t_k }{ \sum_{ k \in \mathbf{S}[t] } |\mathcal{D}_k| }
\end{align}
where $\mathbf{1} \in \mathbb{R}^K$ is the all-one vector and $\odot$ represents the Hadamard product
\EndFor
\State \textbf{Output:} $\mathbf{w}^T$
\label{Alg1:Iteration_End}
\end{algorithmic}
\end{algorithm}

\section{Numerical Results}\label{sec:numerical}
In this section, we carry out simulations to verify the effectiveness of the proposed scheduling policy. Particularly, we evaluate the efficiency of training FL on a support vector machine (SVM) over the MNIST data set, which consists of handwritten numerals, under the proposed ABS policy, and a state-of-the-art approach \cite{YanJiaShi:18}, coined as the MaxPack policy.
The whole data set is participated in 100 non-overlapped portions and assigned to the $K=100$ UEs that are randomly scattered within a disc of a radius 100~m, where the AP locates at the center. There are in total $N=20$ available subchannels, and the transmission on any subchannel is subject to Rayleigh fading with unit mean and the path loss that follows a power law, with the exponent being $\beta=3.5$.
\begin{figure}[t!]
  \centering{}

    {\includegraphics[width=0.95\columnwidth]{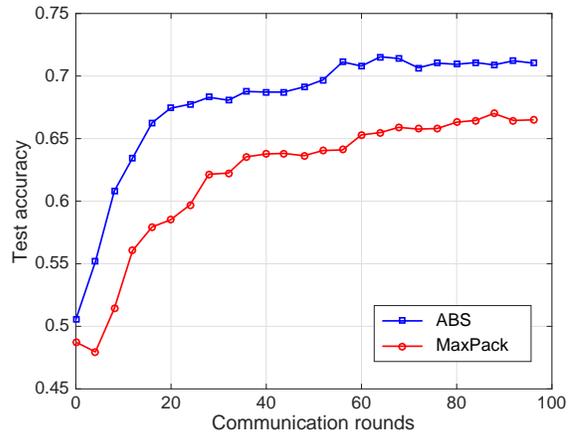}}

  \caption{ Impact of scheduling policies on the convergence of FL in terms accuracy value. }
  \label{fig:Schl_Impact}
\end{figure}


Fig.~\ref{fig:Schl_Impact} plots the test accuracy as a function of the communication rounds. From this figure, we can observe that FL running under the ABS policy is able to attain a marked  improvement in the accuracy value compared to the MaxPack policy. Moreover, the gain achieved by the ABS policy is especially pronounced in the initial stage of the training, e.g., when the communication rounds are less than 40, which demonstrates that the proposed scheme is able to accelerate the convergence of FL and hence improve the learning efficiency.
This gain can be essentially attributed to that while the MaxPack policy exploits the multi-user diversity \cite{TseVis:05} by allocating the best subchannels and power to maximize the total number of receivable updates, the ABS strikes a balance between the aggressive channel use and short term fairness and hence achieves the gradient diversity \cite{YinPanLam:18}.

\section{conclusion}
In this paper, we have proposed a scheduling policy for FL in the context of mobile edge networks. By adopting a metric termed AoU, our scheme is able to account for both the staleness of updates and the instantaneous channel qualities so as to accelerate the convergence rate of FL.

\bibliographystyle{IEEEbib}
\bibliography{bib/StringDefinitions,bib/IEEEabrv,bib/howard_FedSGD_Schl}
\end{document}